\documentclass[aps,prl,twocolumn,superscriptaddress,showpacs,amsmath,amssymb,preprintnumbers]{revtex4}

\usepackage{graphicx,amssymb}
\begin{document}


\title{Magnetic structure of azurite above the 1/3 magnetization plateau}

\author{F. Aimo}
 \affiliation{Laboratoire National des Champs Magn\'{e}tiques
Intenses, LNCMI - CNRS (UPR3228), \\
UJF, UPS and INSA, BP 166, 38042 Grenoble Cedex 9, France}

\author{S. Kr\"{a}mer}
 \affiliation{Laboratoire National des Champs Magn\'{e}tiques
Intenses, LNCMI - CNRS (UPR3228), \\
UJF, UPS and INSA, BP 166, 38042 Grenoble Cedex 9, France}

\author{M. Klanj\v{s}ek}
 \affiliation{Laboratoire National des Champs Magn\'{e}tiques
Intenses, LNCMI - CNRS (UPR3228), \\
UJF, UPS and INSA, BP 166, 38042 Grenoble Cedex 9, France}
 \affiliation{Jo\v{z}ef Stefan Institute, Jamova cesta 39,
1000 Ljubljana, Slovenia}
 \affiliation{EN-FIST Centre of Excellence,
Dunajska 156, 1000 Ljubljana, Slovenia}

\author{M. Horvati\'{c}}
\altaffiliation[Email address: ]{mladen.horvatic@lncmi.cnrs.fr}
\affiliation{Laboratoire National des Champs Magn\'{e}tiques
Intenses, LNCMI - CNRS (UPR3228), \\
UJF, UPS and INSA, BP 166, 38042 Grenoble Cedex 9, France}

\author{C. Berthier}
\affiliation{Laboratoire National des Champs Magn\'{e}tiques
Intenses, LNCMI - CNRS (UPR3228), \\
UJF, UPS and INSA, BP 166, 38042 Grenoble Cedex 9, France}

\date{\today}

\begin{abstract}
The transition from the 1/3 magnetization plateau towards the
saturation magnetization in azurite has been studied by
low-temperature, high-magnetic-field, high-frequency proton
nuclear magnetic resonance (NMR). The observed symmetrical
splitting of the NMR spectra is incompatible with the longitudinal
incommensurate order appearing when the longitudinal correlation
function becomes dominant over the transverse one, which is the
expected framework for the existence of the 2/3 magnetization
plateau. The spectra are rather interpreted in terms of a more
standard transverse antiferromagnetic (canted) order.
\end{abstract}



\maketitle

Since the discovery of the magnetization plateau at 1/3 of the
saturation magnetization in the natural mineral azurite
\cite{Kikuchi05}, Cu$_3$(CO$_3$)$_2$(OH)$_2$, the system has been
intensely studied as a model for the frustrated antiferromagnetic
Heisenberg spin-1/2 chain of ``distorted diamond'' geometry
\cite{Okamoto03}. Its large 1/3 plateau is of pure quantum origin,
as it is based on the spin configuration where two more strongly
coupled ``dimer'' spins are in a singlet state, while the third
``monomer'' spin is fully polarized \cite{Okamoto03,Aimo09}. In
this letter we focus on another interesting subject, the possible
existence of a narrow 2/3 plateau (so far unresolved by the
magnetization measurements \cite{Kikuchi05,KikuchiPTPS05}) in the
middle of the increase of magnetization from the 1/3 plateau
towards the fully polarized system
\cite{Okamoto03,Sakai09,Sakai10}. Unlike the 1/3 plateau, the 2/3
plateau is expected from Oshikawa's criterion \cite{Oshikawa97} to
break the translational symmetry. Within the one-dimensional (1D)
description (i.e., the Tomonaga-Luttinger description), where the
spin-spin correlation functions along the chain exhibit a
power-law behavior, it is associated to a rather exceptional
situation in which the \emph{longitudinal} spin-spin correlation
function becomes dominant over the transverse one due to the
so-called ``$\eta$ inversion'' of the corresponding exponents
($\eta^\parallel < \eta^\perp)$ \cite{Sakai09,Sakai10}. In this
case an incommensurate (IC) longitudinal (i.e., parallel to the
applied field) three-dimensional (3D) order is expected to be
stabilized at low temperature, which can generate a plateau at the
commensurate point. Theoretically it has been shown that the 2/3
plateau may exist in a diamond spin chain
\cite{Okamoto03,Sakai09,Sakai10} for certain values of exchange
couplings $J_1$, $J_2$, $J_3$ lying within the relatively broad
range of values extracted from the experiments
\cite{Kikuchi05,Gu07,Gu_Kikuchi,Rule08,Mikeska08,Kang09,Jeschke11}.
Namely, there is a controversy on these values, which seems to
have converged only very recently \cite{Jeschke11}.

Another, more conventional and thus less interesting situation
appears when the \emph{transverse} correlations are dominant,
leading to a transverse antiferromagnetic (AF) order at low
temperature, meaning that the dimer spins develop a canted
polarization. The order would typically be of the N\'{e}el type
where the direction of the transverse polarization is fixed, but
it may also be a spiral IC order where the transverse polarization
rotates along the chain. In this work we present nuclear magnetic
resonance (NMR) data which are clearly in favor of the N\'{e}el
canted order. We thus \emph{exclude} the more appealing
possibility of the longitudinal IC order leading to a 2/3 plateau.

In order to determine the microscopic nature of the spin structure
of azurite above the 1/3 plateau, we performed a high-frequency
(1.3--1.45\,GHz) proton NMR at very high field (31--34\,T) in the
M9 magnet at LNCMI-Grenoble, using specially designed (``bottom
tuned'') NMR probe for the narrow bore $^3$He refrigerator. We
mention that, in contrast to the previous NMR work at lower field
\cite{Aimo09}, the spin structure in this field range cannot be
directly accessed by the copper NMR, because the signal is lost
(``wiped out'' by shortening of $T_2$) due to enhanced
fluctuations of electronic spins. There is only one proton site in
azurite, which however generates two NMR lines for an arbitrary
direction of the magnetic field $H$, corresponding to two diamond
chains per unit cell which are differently oriented with respect
to $H$ (see Fig.~1 in Ref.~\cite{Aimo09}). When the field is
applied within the $ac$-plane of mirror symmetry, both chains
become equivalent and only one proton NMR line is observed. In
this study the azurite single crystal was oriented with its
$c$-axis approximately parallel to the magnetic field, so that the
field is perpendicular to the chain $b$-axis, and only one proton
NMR line is expected for a homogeneous system.

\begin{figure}[h]
\includegraphics[width=0.9\linewidth,clip]{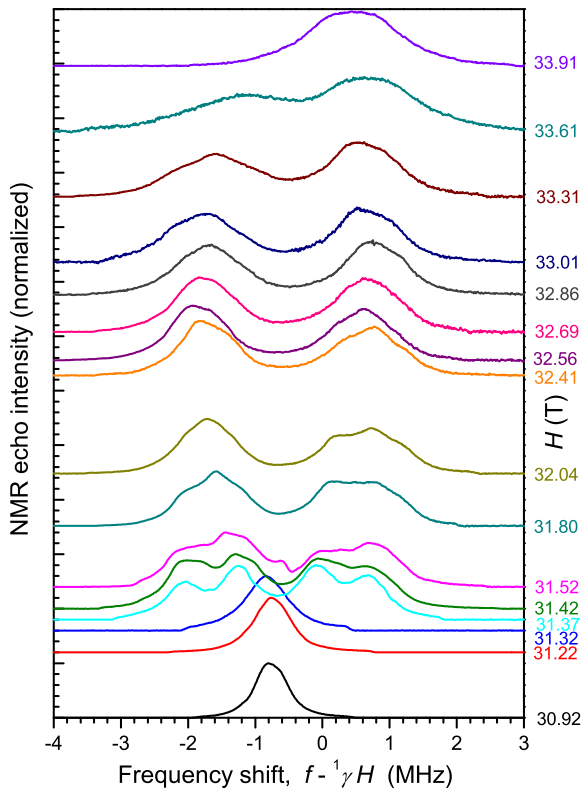}
\caption{\label{fig:one} (color online) High-field proton NMR
spectra of azurite covering the transition from the 1/3 plateau
into the fully polarized system, taken at 0.6\,K. Vertical offset
of the spectra is proportional to the magnetic field, applied
parallel to the $c$-axis. Frequency axis is taken relative to the
proton Larmor frequency defined by its gyromagnetic ratio,
$^1\gamma$~= 42.5774\,MHz/T.}
\end{figure}

Indeed, in the field dependence of proton NMR spectra shown in
Fig.~\ref{fig:one} a single line is observed in the 1/3 plateau,
that is below the second critical field
$H_{\textrm{c2}}$(0.6\,K)~= 31.35\,T. The ``activated'' magnetic
field dependence of the nuclear spin-lattice relaxation rate data
($T_1^{-1}$, not shown here) confirms that this field range
corresponds to a linear closing of the gap at the end of the
plateau, as is also observed by high-field EPR \cite{Ohta03}. At
$H_{\textrm{c2}}$ the NMR line suddenly broadens and then splits
into a nearly symmetrical spectrum. Each side of the spectrum
consists of a pair of lines just above $H_{\textrm{c2}}$. Each
pair merges into a distribution with some structure, and
transforms into a more regular Gaussian lineshape at higher
fields.

\begin{figure}[b!]
\includegraphics[width=0.80\linewidth,clip]{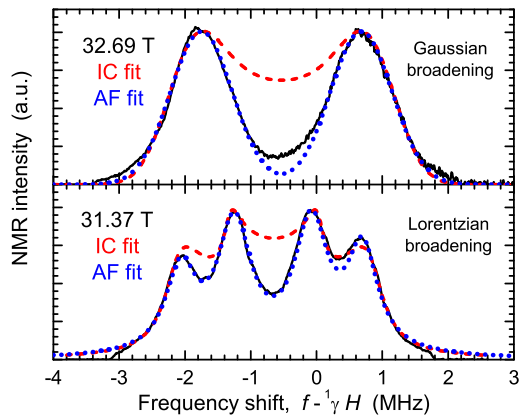}
\caption{\label{fig:two} (color online) Fits of two representative
proton NMR spectra (solid lines) to the lineshapes corresponding
to an AF ordered state (dotted lines) and to a 1D IC state (dashed
lines).}
\end{figure}

As the spectrum is nowhere completely split, it is difficult to
directly distinguish if the lineshape corresponds to a simple
N\'{e}el type splitting into two (or 2$\times$2) somewhat
overlapping lines centered at two frequencies $f_1$ and $f_2$, or
to an IC lineshape. In the latter case, for a 1D sinusoidal
distribution of frequencies spreading between $f_1$ and $f_2$, an
NMR lineshape $I(f)$ corresponds to the density function $D(f)
\propto [(f - f_1)(f_2 - f)]^{-1/2}$ convoluted by some
line-broadening function $G(f)$, $I(f) = D \ast G(f)$. In
Fig.~\ref{fig:two} we have performed the corresponding fits to two
selected spectra presenting the most regular lineshape, one just
above $H_{\textrm{c2}}$ and the other in the middle of the
transition towards the saturation. In the former case the best fit
was achieved using ad hoc Lorentzian line-broadening function,
while the latter spectra were better approximated by a Gaussian
line-broadening. In both cases the fits are in favor of an AF
ordered system, while fits to the lineshape corresponding to an IC
state clearly fail.

\begin{figure}[t!]
\includegraphics[width=1.0\linewidth,clip]{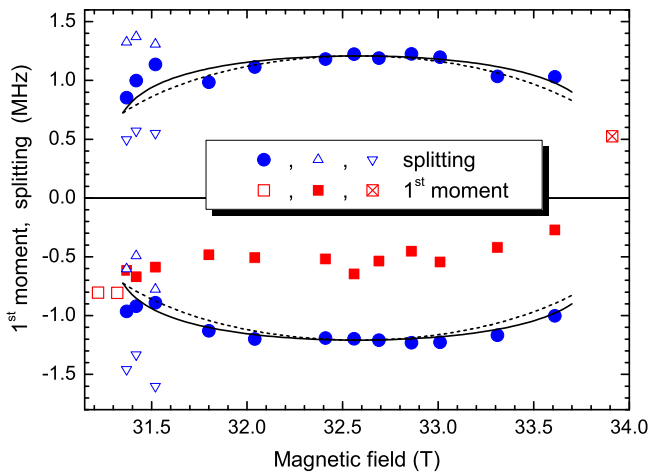}
\caption{\label{fig:three} (color online) Analysis of the spectra
given in Fig.~\ref{fig:one}: The first moment, i.e. the average
lineshift with respect to the Larmor frequency, in the 1/3 plateau
({\scriptsize $\square$}), in the transition region ({\scriptsize
$\blacksquare$}), and close to full polarization ({\scriptsize
$\boxtimes$}). The position of two ({\Large ${\bullet}$}) or four
({\scriptsize $\bigtriangleup$}, {\scriptsize $\bigtriangledown$})
components of the nearly symmetrically split spectra determined
relative to the 1st moment. Close to the 1/3 plateau, where
2$\times$2 components fit is applied, {\Large ${\bullet}$} denotes
the weighted average position of {\scriptsize $\bigtriangleup$}
and {\scriptsize $\bigtriangledown$}. Solid and dashed lines are
predicted field dependence for 1D and 3D system of weakly
interacting dimers, respectively (see the text).}
\end{figure}

Further quantitative analysis of the spectra is given in
Fig.~\ref{fig:three}. We first observe that the first moment of
the complete lineshape, which reflects an average spin
polarization in the system, is barely changing although the
magnetization of the system strongly increases. This can be
attributed to a nearly zero value of the longitudinal coupling
constant A$_{zz}$ of protons for this particular orientation of
the sample, due to the cancellation of the hyperfine
(super-exchange induced) and the direct dipolar component.

In order to quantify the splitting of the spectra shown in
Fig.~\ref{fig:one}, which is apparently nearly field independent,
we have fitted the spectra to two Gaussians (see
Fig.~\ref{fig:two}) and plotted in Fig.~\ref{fig:three} their
position taken with respect to the 1$^{\textrm{st}}$ moment of
each spectrum. For the three spectra close to $H_{\textrm{c2}}$,
where the fit to 2$\times$2 peaks is more appropriate, the
splitting is defined from the position of the weighted average of
the two peaks on each side. In this way we find that the splitting
of the spectra is highly symmetrical and nearly flat with very
weakly expressed maximum at $H_{2/3}$~= 32.6\,T, close to the
midpoint between 1/3 magnetization and saturated magnetization. It
turns out that the observed field dependence of the line splitting
follows precisely the prediction (see Fig.~\ref{fig:three}) for
the transverse staggered spin polarization $S^{\perp}$  in the
simplest model of independent dimers, which has originally been
employed to describe the Bose-Einstein condensation (BEC) of
triplets \cite{Momoi00,comment1}. We observe that close to
$H_{\textrm{c2}}$ one of the Zeeman-split triplet states of a
dimer becomes degenerate with the singlet. Mixing these two states
produces a state where the two spins of a dimer acquire identical
longitudinal polarizations $S^z$ and opposite transverse
polarizations $S^{\perp}$, which depend on the mixing coefficients
\cite{Momoi00}. The magnitudes of the two components are related
by an elliptical dependence:
\begin{equation}\label{eq}
    S^{\perp} = \sqrt{2} \sqrt{1/4^2 - (S^z - 1/4)^2}
\end{equation}
where $S^{\perp}$ has a broad maximum at the point where the spins
are half polarized, $S_{max}^{\perp} = S^{\perp}(S^z = 1/4) =
\sqrt{2}/4$, and develops very rapidly from zero, $S^{\perp}(S^z =
0, 1/2) = 0$, as soon as the system departs from zero or full
polarization. In azurite the polarization of dimers in the 1/3
plateau is found to be 10\,\% \cite{Aimo09}, so that we expect
that at $H_{\textrm{c2}}$ the transverse polarization will develop
starting right away from as much as $S^{\perp}(S^z = 1/20) =
0.6\,S_{max}^{\perp}$. Eq.~\ref{eq} can be used to predict the
field dependence of $S^{\perp}$, if $S^z(H)$ is known from the
experiment or from the theory. High field magnetization data in
azurite \cite{KikuchiPTPS05} show that the increase of
magnetization towards saturation is of the ``$\arcsin$'' type
corresponding to what is expected for a 1D system. Our 1D fit
shown in Fig.~\ref{fig:three} corresponds to the $\arcsin$
dependence defined by the initial point $S^z(H_{\textrm{c2}})$~=
1/20 and the midpoint $S^z(H_{2/3})$~= 1/4. For comparison, we
also show a 3D fit, where the $S^z(H$) dependence is taken to be
linear, passing through the same points. The 1D fit is excellent,
in particular when we take into account that the only fitting
parameters are the overall scaling defined by the observed maximum
splitting of $\Delta f_{max}$~=  1.2\,MHz and the corresponding
field value $H_{2/3}$. The 3D fit is very close, and the
experimental data do not really allow us to distinguish between
the two fits. The fits confirm the description in terms of the
transverse polarization, whose main characteristics is that strong
increase of magnetization is accompanied by nearly field
independent splitting of the spectra. We note that this particular
type of behavior is \emph{not} expected in a description in terms
of an IC modulation of the longitudinal polarization, which
provides another argument against this latter interpretation.

\begin{figure}[b!]
\includegraphics[width=0.90\linewidth,clip]{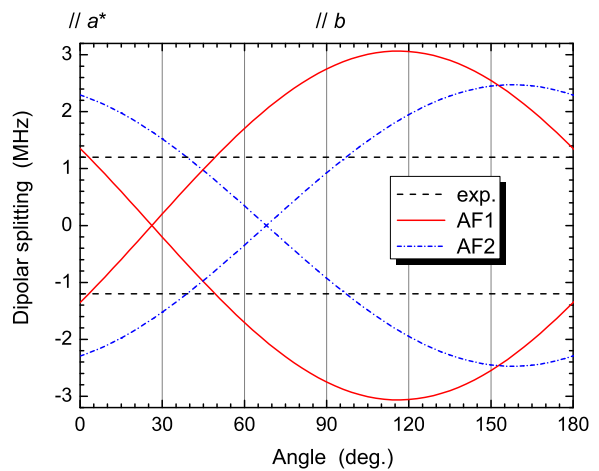}
\caption{\label{fig:four} (color online) Experimental value of the
line splitting (dashed line) compared to the predicted dipolar
splitting for two possible AF orders (solid and dash-dotted line,
as a function of the orientation of the ordered moments in the
transverse plane.}
\end{figure}

In the following we proceed by a quantitative analysis of the
experimental value ($\Delta f_{max}$) of the splitting. A
transverse component of the spin polarization is shifting the NMR
line through the off-diagonal (A$_{zx}$) elements of the coupling
tensor. In general both the direct dipolar field, as well as the
hyperfine field are expected to contribute to the proton coupling
tensor. However, the latter mechanism is based on an $s$-type wave
function at the proton site, it is thus isotropic and cannot
generate A$_{zx}$ elements. This means that the NMR splitting due
to the transverse staggered $S^{\perp}$ is of pure dipolar origin
and can be calculated for a given spin order from the known
crystallographic structure \cite{Belokoneva01,Rule11,comment2}. In
Fig.~\ref{fig:four} we show so predicted NMR linesplitting
generated by $\pm S_{max}^{\perp} = \pm \sqrt{2}/4$ on dimer spins
as a function of their direction in the plane perpendicular to the
field applied along the $c$-axis. There are two possibilities for
the stacking of $S^{\perp}$ (AF1 \cite{comment3} and AF2) as
regards the relative phase (0 or $\pi$) of an AF spin arrangement
in two chains in a unit cell. Comparing the prediction to the
experimental value of $\pm$1.2\,MHz, we observe that this is very
close to the predicted $\pm$1.35\,MHz for the AF1 type order with
$S^{\perp}$ along the $a^* \cong a$-axis. This particular
direction agrees with the easy magnetization axis (perpendicular
to the chains), as defined by the low-field magnetization
measurements \cite{Kikuchi05}. However, the magnetization data
correspond to the initial polarization of the monomer spins, while
in the high-field measurement these spins are saturated and thus
inert, and the active sites are dimers. Unfortunately, there is no
direct information on the ``local'' easy axis of dimers at high
field. In general, azurite presents quite strong anisotropy of its
magnetic properties, which has not yet been understood
\cite{Kikuchi05,Sakai10,Gu07,Rule11}.

One should remark that the predicted splitting from Eq.~\ref{eq}
corresponds to the maximum expected $S^{\perp}$, which is then to
some extent reduced by thermal and quantum fluctuations. According
to the complete mean field solution for the 3D interacting dimers
\cite{Tachiki70}, the thermal effects are negligible below $0.4\,
T_c^{max}$, where $T_c^{max}$ is the maximum critical temperature.
From measurements taken at 1.2\,K we know that $T_c^{max}$ is well
above this temperature, meaning that the low temperature limit is
fulfilled for our 0.6\,K spectra. The effect of quantum
fluctuations depends strongly on dimensionality and can be very
important for purely 1D systems. For example, in the
CuBr$_4$(C$_5$H$_{12}$N)$_2$ (BPCB) spin ladder, where the
effective 3D coupling is $\approx$50 times smaller than the
inter-dimer interactions, the reduction of the transverse moment
is $\approx$50\,\% \cite{Klanjsek08,Bouillot}. In azurite the 3D
exchange couplings \cite{Jeschke11} are estimated to be of the
order of a few K, that is (only) one order of magnitude smaller
than the (effective) 1D couplings. At high magnetic field the 1D
Hamiltonian of azurite can be further reduced to the effective
XXZ, $S = 1/2$ spin chain Hamiltonian describing only the dimer
spins \cite{Honecker01,Honecker11}. The corresponding couplings
are of the order of 1\,K, of the same size as the ordering
temperature, so that we can suspect that residual 3D couplings are
of comparable size. The 3D character of azurite at high field is
thus probably important, making the effect of quantum fluctuations
relatively weak. For example, comparing our prediction for
$S^{\perp}$ along the $a^*$-axis to the experimental value we get
the reduction of only $\approx$10\,\%. We also recall that this
description does not take into account the anisotropy terms of
azurite's Hamiltonian \cite{Kikuchi05,Sakai10,Gu07,Rule11}, which
might also significantly modify the observed spin polarizations.

Finally, within the effective XXZ Hamiltonian the anisotropy of
the numerically optimized \cite{Honecker11} effective coupling is
found to be $J_{xy}/J_z$~= 2.2, very close to the value of 2
obtained when the same description is applied to the spin ladder
in the strong coupling regime \cite{Bouillot}. For this latter
system we know that in the 3D ordered phase the spins are canted
and that no plateau is expected, in agreement with our NMR
results.

In conclusion, the high-field proton NMR spectra of azurite have
been analyzed to determine which type of spin order is realized
between the 1/3 magnetization plateau and the full polarization of
the system. Both the observed line shape and the field dependence
of the splitting of the spectra clearly correspond to the
``conventional'' \emph{transverse} (canted) N\'{e}el order. They
are both incompatible with the longitudinal IC order, meaning that
there is no 2/3 plateau in azurite, and that the corresponding
``more interesting'' physics does not apply to this system. These
conclusions are supported by the recent theoretical estimates of
the effective Hamiltonian for this system \cite{Honecker11}.

\begin{acknowledgments}

We acknowledge fruitful discussions with B. Grenier, A. Honecker,
H. Ohta, and T. Sakai. Part of this work has been supported by the
French ANR project NEMSICOM and by European Commission through the
Transnational Access Specific Support Action (Contract No.
RITA-CT-2003-505474), by the EuroMagNET network (Contract No.
RII3-CT-2004-506239), and by the ARRS project No. J1-2118.

\end{acknowledgments}


\begin{references}
\bibitem{comment1}
A BEC cannot apply to azurite because of its magnetic
anisotropy (discussed later on).
\bibitem{comment2}
The longitudinal spin polarization does not need to be taken into
account because it is homogeneous over all the dimer spins and
therefore cannot produce any linesplitting.
\bibitem{comment3}
In AF1 the dimers related by the (0, 1/2, -1/2) translation vector
have the same spin polarization, as in Fig. 6 of
Ref.~\cite{Rule11}.
\end{references}


\end{document}